\documentclass[pdflatex,sn-nature]{sn-jnl}
\usepackage{aas_macros}
\usepackage{amsmath,amssymb,mathtools,comment}

\def\exok1{\texttt{Exo\_k}}

\jyear{2022}%

\begin{document}

\title[Cool runaway greenhouse]{A cool runaway greenhouse without surface magma ocean}


\author*[1]{\fnm{Franck} \sur{ Selsis}}\email{franck.selsis@u-bordeaux.fr}

\author[1]{\fnm{Jérémy} \sur{Leconte}} \email{jeremy.leconte@u-bordeaux.fr}

\author[2]{\fnm{Martin} \sur{Turbet}} \email{martin.turbet@lmd.ipsl.fr}
\author[3]{\fnm{Guillaume} \sur{Chaverot}} \email{guillaume.chaverot@unige.ch}
\author[3,4]{\fnm{Émeline} \sur{Bolmont}} \email{emeline.bolmont@unige.ch}
\affil*[1]{Laboratoire d'astrophysique de Bordeaux, Univ. Bordeaux, CNRS, Pessac, France}

\affil[2]{Laboratoire de M\'et\'eorologie Dynamique/IPSL, CNRS, Sorbonne Universit\'e, \'Ecole Normale Sup\'erieure, PSL Research University, \'Ecole Polytechnique, 75005 Paris, France}

\affil[3]{Observatoire Astronomique de l’Université de Genève, Chemin Pegasi 51, Sauverny 1290, Switzerland}
\affil[4]{Center for Life in the Universe, Faculty of Science, University of Geneva, Geneva, Switzerland}


\abstract{Water vapour atmospheres with content equivalent to the Earth's oceans, resulting from impacts\cite{2014RSPTA.37230172S} or a high insolation\cite{1969JAtS...26.1191I,1988Icar...74..472K}, were found to yield a surface magma ocean\cite{2013Natur.497..607H, 2013JGRE..118.1155L}. This was, however, a consequence of assuming a fully convective structure\cite{1969JAtS...26.1191I,1988Icar...74..472K,1993Icar..101..108K, 2007A&A...476.1373S, 2013Natur.497..607H, 2013JGRE..118.1155L, 2013NatGe...6..661G, 2013ApJ...765..131K, 2016ApJ...822...24D,2021PSJ.....2..207G}. Here we report, using a consistent climate model, that pure steam atmospheres are commonly shaped by radiative layers, making their thermal structure strongly dependent on the stellar spectrum and internal heat flow. The surface is cooler when an adiabatic profile is not imposed: melting Earth's crust requires an insolation several times higher than today, which will not happen during the main-sequence of the Sun. Venus' surface can solidify before the steam atmosphere escapes, which is opposite to previous works\cite{2013Natur.497..607H, 2013JGRE..118.1155L}. Around the reddest stars ($T_{eff}<3000$~K), surface magma oceans cannot form by stellar forcing alone, whatever the water content. These findings affect observable signatures of steam atmospheres and exoplanet mass-radius relationships, drastically changing current constraints on the water content of Trappist-1 planets. Unlike adiabatic structures, radiative-convective profiles are sensitive to opacities. New measurements of poorly constrained high-pressure opacities, in particular far from the H$_2$O absorption bands, are thus necessary to refine models of steam atmospheres, which are important stages in terrestrial planet evolution.}

\keywords{Runaway greenhouse, magma ocean, Earth, Venus}



\maketitle

\renewcommand\thefigure{Figure~\arabic{figure}}  
\renewcommand{\figurename}{}
\section{Main article}\label{sec:main}
Modeling the steam atmosphere resulting from the vaporization of the Earth oceans (270~bars) is challenging. If a few consistent 1D \cite{2019ApJ...886..140K} and 3D simulations have been achieved up to 10-30~bars\cite{2021Natur.598..276T}, most models relied on the hypothesis that steam atmospheres are fully convective below their photosphere. This assumption allowed to 1) building the thermal structure from a surface temperature $T_\textrm{surf}$ using dry and saturated adiabats and 2) determining the corresponding incoming stellar radiation (ISR) by \textit{inverse climate modeling}, i.e. by insuring a radiative balance at the top of the atmosphere based on the computed outgoing thermal radiation (OTR) and Bond albedo A. In this framework, the OTR decreases slowly from the Nakajima-Simpson limit\cite{1992JAtS...49.2256N} of $\sim 280-290$~Wm$^{-2}$ (depending on the H$_2$O mixing ratio\cite{2022A&A...658A..40C}, clouds\cite{2021Natur.598..276T} and gravity\cite{2014ApJ...787L..29K}) to a value of $\sim 274$~Wm$^{-2}$ over a broad range of surface temperatures: from about 300~K to 1700~K. Above 1700~K the OTR increases again with T$_\textrm{surf}$. Therefore, once the absorbed flux exceeds the runaway threshold, the surface rapidly evolves from \textit{habitable} conditions into a magma ocean. Conversely, when a hot and cooling steam atmosphere above a magma ocean reaches an OTR of 274~Wm$^{-2}$ and a T$_\textrm{surf}$ of 1700~K, it enters a fast cooling phase, leading to solidification of the crust and condensation into an ocean.

However, as shown in the section \textit{Most steam atmospheres are not fully convective} (\ref{subsec:method1}), balancing the radiative fluxes at the top of a convective atmosphere, as done in \textit{inverse climate modeling}, does not insure that all layers are at thermal equilibrium as they should. Additionnaly, we found that they are not in most of the parameter space we explored: surface pressure $p_\textrm{surf}$ between 0.1 and 270~bars, stellar effective temperature $T_{eff}$ between 2600 and 7000~K, OTR between 274 and $10^5$~Wm$^{-2}$ and internal heat flux between 0 and 1000~Wm$^{-2}$. It is therefore necessary to self-consistently calculate the atmospheric thermal profile, without imposing convection, following the tracks of pioneering works on runaway atmospheres \cite{1984E&PSL..68....1W} and formation/condensation of post-accretion atmospheres \cite{1988JAtS...45.3081A}. This is required to reassess results obtained assuming fully convective profiles on the divergent fate of Venus and the Earth\cite{1988Icar...74..472K,2013Natur.497..607H}, the circumstellar region where surface liquid water can be stable\cite{1993Icar..101..108K,2013ApJ...765..131K} and the occurrence of surface magma oceans\cite{2013Natur.497..607H,2013JGRE..118.1155L,2015ApJ...806..216H,2017JGRE..122.1458S,2021JGRE..12606711L}.

\begin{figure*}[htb]
\centering
\includegraphics[width=\linewidth]{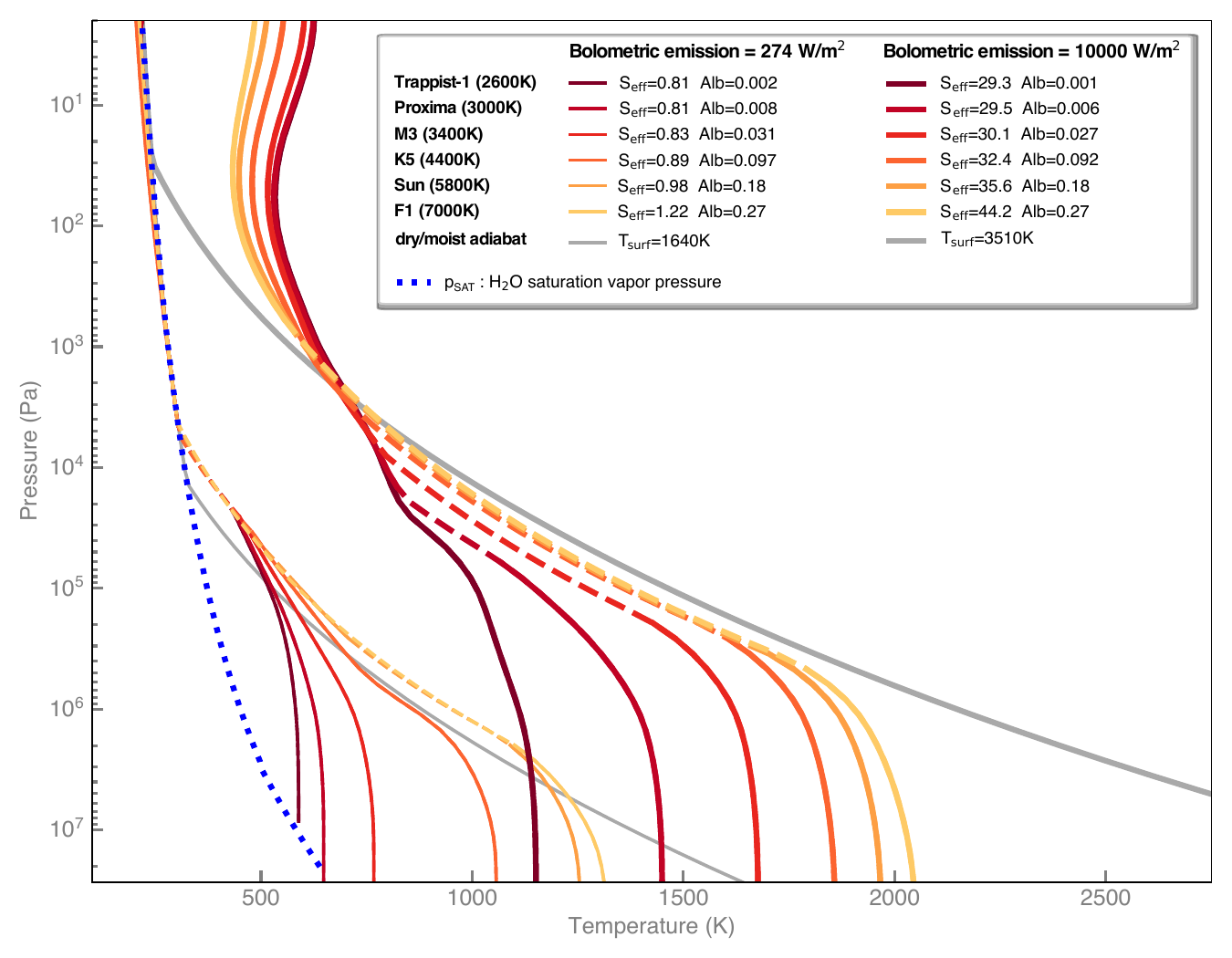}
\caption{
\textbf{Steam atmosphere thermal profiles}. Each group of five profiles shares the same bolometric emission (274 and 10,000~Wm$^{-2}$, thin and thick lines respectively). The profiles computed for different stellar spectra are compared with convective profiles of same emission. For each computed profile, the Bond albedo and instellation are given (S$_\textrm{eff}=1$ corresponds to Earth's present insolation: 341.5~Wm$^{-2}$).
Dashed lines indicate dry convection. }
\label{fig:Fig1_profils_same_OTR}
\end{figure*}

To calculate the vertical structure consistently with the incoming stellar radiation and underneath heat flow, we developed an evolution code within the \texttt{Exo\_k} library \cite{2021A&A...645A..20L} specifically to handle the coexistence of very large evolution timescales at high depth with very short ones in the upper atmosphere. With \texttt{Exo\_k} (described in details in the Methods), we computed atmospheric profiles in equilibrium with different stellar fluxes/spectra and compared them in 
\ref{fig:Fig1_profils_same_OTR} with convective profiles that share the same bolometric emission. We considered a sample of stars from the ultracool dwarf Trappist-1 (hereafter T-1, $T_{\mathrm{eff}}=2600$~K) to a F1-type star ($T_{\mathrm{eff}}=7000$~K). 
Profiles with an OTR of 274~Wm$^{-2}$ are at the transition between the hot regime where the OTR is sensitive to $T_\textrm{surf}$ and the unstable plateau of the runaway or condensation (depending whether the evolution proceeds toward increasing or decreasing surface temperature). In the full-convection scenario, the planet exits runaway mostly when the dry troposphere reaches and warms the photosphere, increasing the OTR \cite{boukrouche_beyond_2021}. For a given OTR, all profiles cross within the photospheric region, where the temperature controls the emission, and which lies within saturated layers for an OTR of 274~Wm$^{-2}$.

Here, we identify two processes keeping steam atmospheres mostly stable against convection and allowing the atmosphere to equilibrate at lower surface temperatures.
First, a radiative zone can form above the surface depending on the surface pressure and the internal heat flux and can potentially extend to the whole atmosphere, depending on the stellar type and incoming flux (see \ref{fig:ED_Fig2_all_profiles}). This results from the vanishing stellar flux at depth, coupled with an upward radiative transport whose efficiency increases with temperature. Indeed, the thermal gradient needed to radiate a given thermal flux decreases with increasing temperature at fixed opacity, as the radiative diffusivity scales as d$B$/d$T$ $\sim$ $T^3$, where $B$ is Planck's spectral radiance. This effect is well known in hot Jupiter atmospheres where radiative layers extend to pressures as high as 100-1000~bars depending on instellation and heat flow \cite{2019ApJ...884L...6T}. Second, the absorption of stellar light directly warms the middle atmosphere so that the temperature of the photosphere is not set only by either the saturation pressure $p_{sat}(T)$ or an adiabat. This reduces further the contrast between the photospheric and surface temperature for a given OTR. At higher ISR, condensation is totally inhibited making some profiles entirely radiative as can be seen in \ref{fig:ED_Fig2_all_profiles}. The departure from purely convective profiles is not restricted to cases with high pressures blocking the penetration of the stellar radiation toward the surface. Indeed, the intensity and spectral distribution of the stellar flux influence the thermal structure of steam atmospheres whatever the surface pressure as illustrated in \ref{fig:ED_Fig6_Psurf_Fint_EarlyVenus_T1}.

 Compared with computed equilibrium profiles, convective structures dramatically overestimate $T_\textrm{surf}$ by 400~K (Sun) to 1100~K (T-1) at the exit of the runaway greenhouse (OTR=274~Wm$^{-2}$), and by 1500~K (Sun) to 2300~K (T-1) for an OTR of 10,000~Wm$^{-2}$ for the Earth's surface water reservoir (270~bars).

\begin{figure}[ht]
\centering
\includegraphics[width=\linewidth]{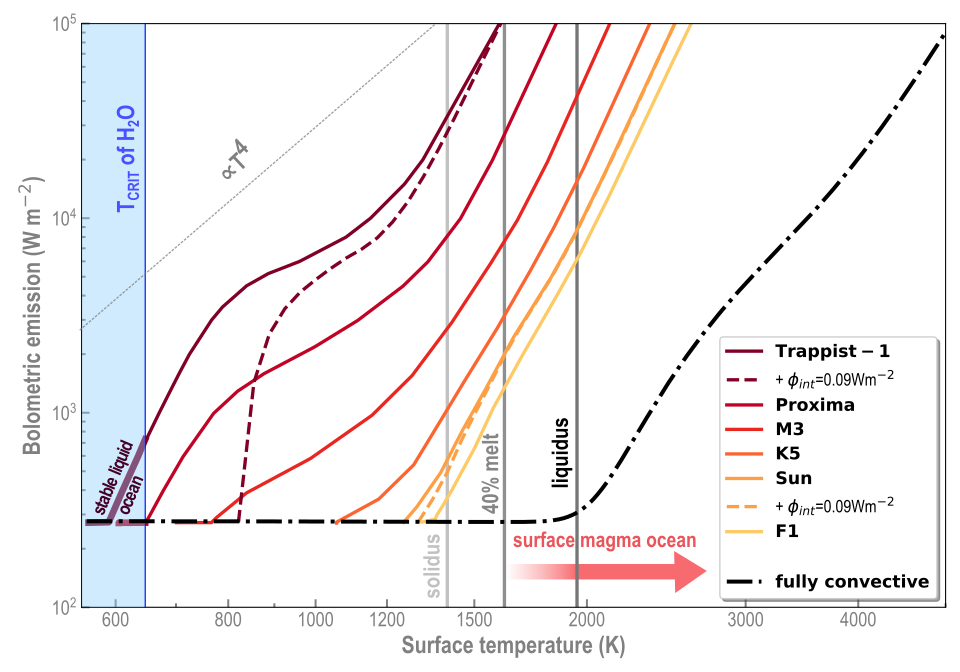}
\caption{
\textbf{Outgoing thermal emission as a function of surface temperature}. The dashed-dotted line is the classic runaway curve obtained with fully convective profiles, independently from the stellar radiation spectrum and internal flux. With self-consistent profiles, this function depends on the type of the host star and the internal flux $\phi_\textrm{int}$. Dashed curves are obtained with Earth geothermal flux. For T-1 and $\phi_\textrm{int}<0.01$~Wm$^{-2}$, the runaway greenhouse can end without full vaporization of the ocean. Low-pressure liquidus, solidus and \textit{melting} (melt fraction = 40\%) temperatures are indicated for an Earth mantle composition. Surface pressure is 270~bars except in the blue area where $T_\textrm{surf}$ is below the critical temperature of H$_2$O and $p_\textrm{surf}$ is given by the saturation vapor pressure of H$_2$O. These curves are based on the profiles shown in \ref{fig:ED_Fig2_all_profiles}}
\label{fig:Fig2_runaway_exit}
\end{figure}

When using convective profiles, runaway extends to surface-melting temperatures \cite{2013ApJ...765..131K,2014ApJ...787L..29K,2013NatGe...6..661G,2015AsBio..15..362G,2015ApJ...806..216H,2019ApJ...875...11N,2019A&A...628A..12T,boukrouche_beyond_2021}. However, in all converged cases and as seen in \ref{fig:Fig2_runaway_exit}, planetary cooling becomes sensitive to $T_\textrm{surf}$ at lower temperature: under solar forcing, the runaway phase ends at a surface temperature around 1250~K. For a typical mantle composition, the low-pressure solidus and liquidus are found at 1400 and 1950~K, respectively \cite{2018NatGe..11..139A}. At 1620~K the melt fraction reaches 40\% and dominates the rheology \cite{Salvador_Samuel22}. This \textit{melting} temperature requires S$_\textrm{eff}>7$ under solar forcing. 

Neither the Earth or Venus have been or will be exposed to such high ISR during the main sequence of the Sun. As a result, a surface magma ocean can be sustained only as long as the interior is able to output a sufficient geothermal flux, either from radiogenic heating or leftover accretion energy. We thus expect that surface magma oceans solidify more rapidly than predicted by previous calculations.
With the insolation of Venus 4.5 Gyrs ago, the crust would solidify for internal internal heat flows below 2, 10 and 100 Wm$^{-2}$ for p$_\textrm{surf}$ = 270, 100 and 10~bars of steam respectively. The post-accretion atmosphere of Venus must have consisted in other species than H$_2$O so we should not draw conclusion from these numbers, but the standard scenarios for Venus early history will have to be revisited with consistent profiles. In particular, surface solidification may not require the loss of water through atmospheric escape in contrast with previous findings\cite{2013Natur.497..607H, 2013JGRE..118.1155L}.

We find a stable steam atmosphere for a solar flux as low as S$_\textrm{eff}=0.98$, a consequence of a known bistability \cite{2013NatGe...6..661G, 2021Natur.598..276T}: a higher ISR is required to vaporize an ocean (S$_\textrm{eff}=1.11$ \cite{2013Natur.504..268L}) than to condense a steam atmosphere (S$_\textrm{eff}=0.92$ \cite{2021Natur.598..276T}). The difference between the 3D simulation including clouds and our cloud-free 1D model is attributed to the greenhouse warming by nightside clouds.

Profiles computed for the star T-1 are fully radiative except below the saturated layers, where dry convection is locally triggered by rain evaporation cooling. The  steam atmosphere is more opaque to its red radiation and only 0.001\% of the incident flux reaches 100~bars, resulting in low surface temperatures. The overestimation of $T_\textrm{surf}$ by convective profiles is considerable: 1200~K for S$_\textrm{eff}=0.8$, just above the runaway threshold, and 2700~K for S$_\textrm{eff}=58$. No planet is close enough to T-1 to be exposed to an instellation large enough to melt the surface under a steam atmosphere, even during the pre-main-sequence (PMS) phase. This has important consequences as it weakens the coupling between reservoirs in the atmosphere and the mantle. Losses to space during PMS can exhaust the steam atmosphere while keeping intact the internal reservoir. In previous models assuming a steam atmosphere in equilibrium with a surface magma ocean, atmospheric escape draws water from the mantle and can desiccate the whole planet \cite{2013Natur.497..607H}. 

This work shows that ultracool dwarfs ($T_{eff} \leq 2600$~K) can theoretically host a hot liquid ocean under a dense steam atmosphere despite an ISR beyond the runaway limit. However, this case remains somehow academic because this requires a high surface pressure ($>$100~bars), which produces a thick blanket for the internal heat. A fraction of Earth's geothermal flux is then sufficient to heat the surface above the critical temperature of H$_2$O (see \ref{fig:Fig2_runaway_exit} and \ref{fig:ED_Fig6_Psurf_Fint_EarlyVenus_T1}).

\begin{figure}[tb]
\centering
\includegraphics[width=1.0\linewidth]{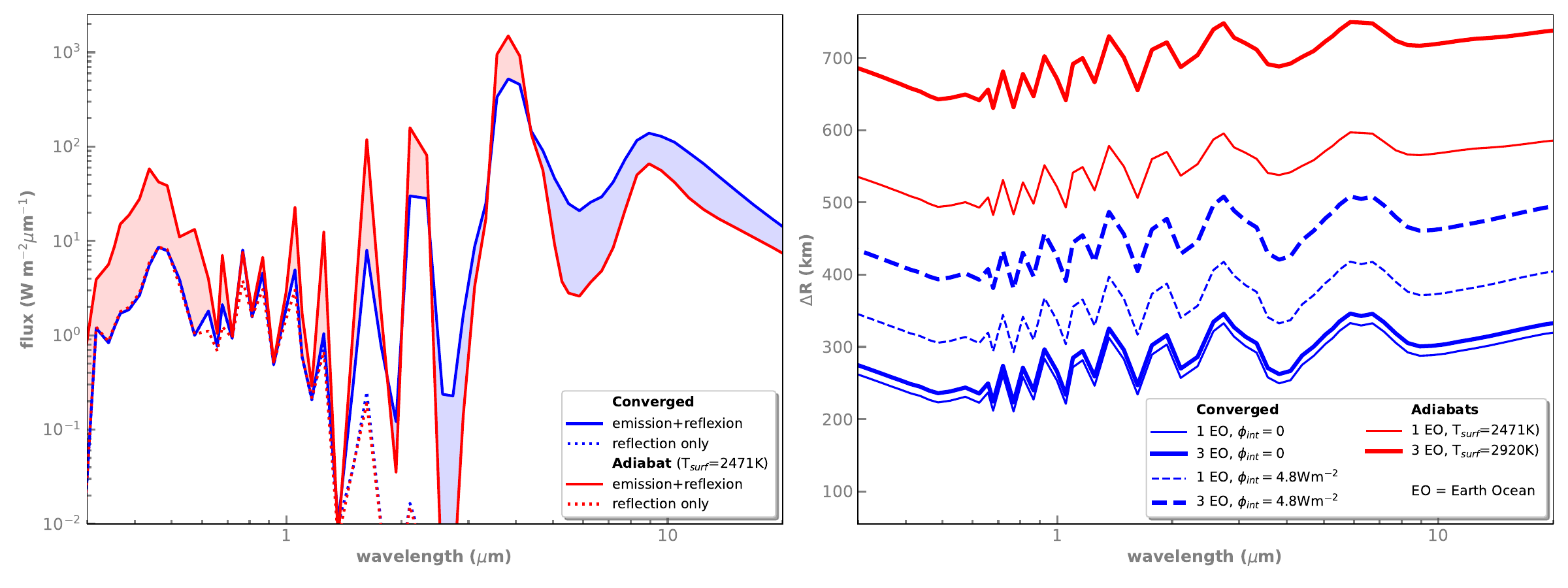}
\caption{\textbf{Spectra of Trappist-1b with a steam atmosphere}. Left: Top-of-the-atmosphere emission and reflection spectra of T-1b with a 270-bars steam atmosphere. Spectra are computed with a converged profile (blue) and with a $T_\textrm{surf}=2410$~K convective profile satisfying OLR = ISR(1-A) (red). Right: transmission spectra given as apparent radius variations. Converged profiles are in blue and given for either 1 or 3 Earth oceans and $\phi_\textrm{int}=0$ or 5~Wm$^{-2}$ (a range of estimated tidal dissipations). 
}
\label{fig:Fig3_spectra_T1b}
\end{figure}

The interpretation of planetary spectra and transit lightcurves differs whether one assumes convective structures or calculates actual profiles. This applies to all exoplanets beyond the runaway limit that could possess a significant water reservoir but let us illustrate this point with the three innermost planets of Trappist-1, which offer incomparable observation conditions for planets with this size and instellation. Occultations of planets b and c and transits of all T-1 planets are observed during the first cycle of the James Webb Space Telescope. 

At a given ISR, the higher temperatures found with convective profiles at pressures higher than $\sim 100$~mbar are responsible for the two main differences, shown in \ref{fig:Fig3_spectra_T1b} for the conditions of T-1b. First, it produces a strong emission between the water bands below 2~$\mu$m and in particular, in the visible, where it exceeds the flux reflected by the planet. With converged profiles this shortwave emission is basically absent while the stellar flux absorbed in the upper atmosphere is re-emitted at low temperature, enhancing the longwave emission. Due to larger scale heights at higher temperature, the altitude of the highest opaque layers is increased in convective profiles.

As a consequence, we can see in \ref{fig:Fig3_spectra_T1b} that convective profiles shift the transmission spectra by several hundreds kilometers for the same surface pressure. This overestimation of the runaway greenhouse radius inflation effect \cite{2019A&A...628A..12T} by the steam atmosphere yielded an underestimation of the maximum water content of T-1b, c and d \cite{2020A&A...638A..41T,2021PSJ.....2....1A,2021A&A...647A..53A,2021ApJ...922L...4D}. 
With \textit{inverse climate modeling}, the internal heat flux only affects the value of $T_\textrm{surf}$ corresponding to top-of-the-atmosphere radiative balance. Outside the runaway OTR plateau, only internal heat flows reaching a significant fraction of the instellation affect atmospheric and spectral properties. With actual profiles, the deep atmosphere that is deprived of stellar light is very sensitive to the internal heat flux, which becomes a parameter controlling the planet radius.
For T-1b, with one and three Earth oceans, the convective hypothesis causes an overestimation of the radius, in all observation bands, by 300~km (4.2\%) and 400~km (5.6\%), respectively. Accounting for an internal heat flow due to tidal dissipation and estimated at 5~Wm$^{-2}$\cite{2018A&A...612A..86T,2018ApJ...857..142M,2018A&A...613A..37B,2018MNRAS.476.5032P,2020A&A...644A.165B} adds $\sim 100$~km (1.4\%) to the radius. The current 1-$\sigma$ relative precision on the radius of T-1b is 1\%\cite{2021PSJ.....2....1A}.

Our calculations thus have important consequences for the interpretation of the exoplanet radius gap \cite{2017AJ....154..109F} and of the true nature of sub-Neptune size planets \cite{2021JGRE..12606639B,2022Sci...377.1211L}. Indeed, we show that water-rich atmospheres are significantly less extended than previously calculated\cite{2020A&A...638A..41T,2020ApJ...896L..22M,2021ApJ...914...84A}, which challenges the interpretation of the bulk population of sub-Neptune size planets as being water worlds.

\textit{Inverse climate modeling} based on the full convection assumption is independent of the H$_2$O opacity at high pressure, sensitive only to the Bond albedo and OTR controled by low-pressure layers. Profiles consistently at equilibrium with the incoming stellar radiation and internal heat flow prove dependent on opacities at all pressures. These opacities control conditions in the deep atmosphere and at the surface (see \ref{sub:opacity} in Methods). The current lack of experimental constraints, in particular for H$_2$O high-pressure far-wing opacities, thus becomes a critical issue. Our conclusions that radiative layers must be considered and that magma oceans do not result from stellar forcing alone around solar and subsolar stars are robust against orders of magnitude changes in these high-pressures opacities. This is mainly due to to the strong temperature feedback that makes radiative transport more efficient at high temperature. Nevertheless, computed thermal profiles remain quantitatively uncertain and we hope this study will motivate measurements of these data, needed to better model the properties and evolution of a broad range of planets, including our own.

\section{Methods}\label{sec:methods}
\renewcommand\thefigure{Extended Data Figure~\arabic{figure}}    
\setcounter{figure}{0}

\subsection{Most steam atmospheres are not fully convective} \label{subsec:method1}

\begin{figure*}[b]
\centering
\includegraphics[width=\linewidth]{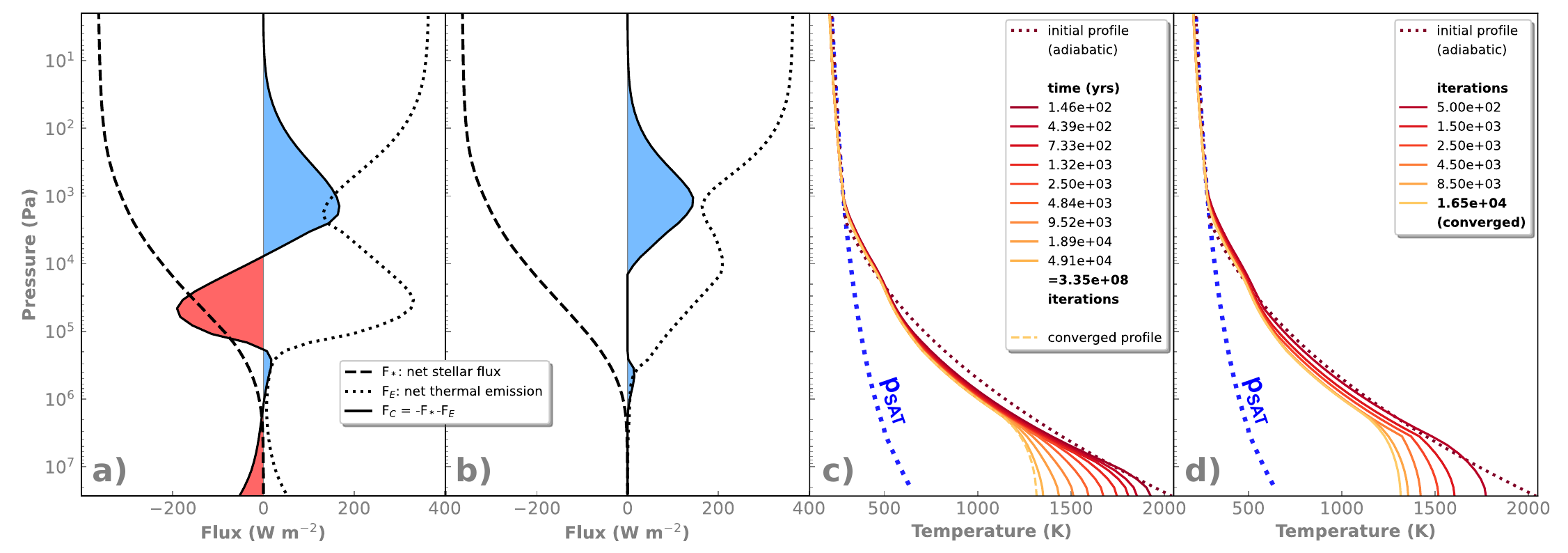}
\caption{\textbf{Converged versus fully convective profiles.} The steam atmosphere is modeled here for a solar flux of 445.7~Wm$^{-2}$ ($S_\textrm{eff}=1.3$), no internal heat flux, a 1g gravity and a 270~bars surface pressure (1 vaporized Earth ocean). \textbf{Panel a}: Net fluxes computed for a ``convective'' P-T profile used as the initial structure in panels c and d. \textbf{Panel b}: Net fluxes at equilibrium. Blue (resp. red) area indicate a downward (resp. upward) net radiative flux. Convection transporting energy upward can only balance a downward radiative flux. Red area therefore indicate a departure from thermal equilibrum.  
\textbf{Panel c)} Evolution from the initial dashed ``convective'' profile towards the converged state, computed with the  \texttt{Exo\_k} suite\cite{2021A&A...645A..20L} evolution package. It takes more than 40,000 yrs to reach the converged state and more than 280 millions steps and 40~hrs of CPU time. \textbf{Panel d)} Evolution computed with an acceleration mode (see Methods) of \texttt{Exo\_k}, in 165,000 steps and less than 1~min of CPU time. In the acceleration mode, iterations and intermediate profiles do not correspond to physical times and structures.}
\label{fig:ED_Fig1_evol_acc}
\end{figure*}

Although a convective structure is well established at the onset of the runaway\cite{2013Natur.504..268L}, a simple test demonstrates that convection cannot be maintained by stellar radiation beyond the runaway threshold. At equilibrium, net ($\uparrow - \downarrow$) energy fluxes must satisfy at each level:
$$
\label{eq:fluxes}
 F_*(p) + F_\textrm{E}(p) + F_\textrm{C}(p) = \phi_\textrm{Int}
$$
where $p$ is the pressure level, $F_*$ is the net stellar flux, $F_\textrm{E}$ is the net thermal emission flux and $\phi_\textrm{int}$ is the internal heat flux. As convection can only transport heat upward, this convention implies that the convective flux, $F_{\textit{C}}(p)$, must be positive in a physically consistent model. \ref{fig:ED_Fig1_evol_acc}.a shows the net radiative thermal and stellar fluxes computed for a convective steam atmosphere with $T_\textrm{surf}=2000~$K and for an ISR insuring a null net radiative flux at the top of the atmosphere. The profile constructed with adiabats is not at local equilibrium and balancing the net radiative flux would require an unphysical \textit{downward} convective flux. The solar flux reaching the lower atmosphere is insufficient to compensate for the strong thermal cooling resulting from the assumed convective gradient: only 12, 0.3 and 0.025~Wm$^{-2}$ of solar radiation reach a depth of 10 and 100 and 270~bars, respectively.  This net flux inconsistency, illustrated here on a specific profile, is systematically observed for $T_\textrm{surf}>700K$ for $P_\textrm{surf}$=270~bars, a solar spectrum, and Earth gravity.

\subsection{The model} \label{subsec:methods/model}

To address this issue we use the tools from the \texttt{Exo\_k} evolution package.  Panels b and c of \ref{fig:ED_Fig1_evol_acc} show the evolution from an initial convective profile at top-of-the-atmosphere radiative equilibrium towards a structure in complete equilibrium (at all pressure levels). In the converged case, $T_\textrm{surf}$ is no less than 700~K cooler than in its assumed convective counterpart. The advantage of \texttt{Exo\_k} is that it uses several physical-acceleration techniques able to accurately reach a radiative-convective equilibrium within short CPU times.  

The \texttt{Exo\_k} library\cite{2021A&A...645A..20L} includes a class dedicated to model self consistently the evolution of planetary atmospheres.
It accounts for radiation, dry convection, turbulent diffusion, and moist processes (condensation, moist convection, and precipitations) for any number of species. We provide a brief description of the salient features of the model below. More details can be found in the online documentation at \url{http://perso.astrophy.u-bordeaux.fr/~jleconte/exo_k-doc/index.html}.\\

\paragraph{Radiative transfer}
The radiative transfer is computed using a 2-stream approximation\cite{TMA89}. The opacities of the radiative layers are computed using the \texttt{Exo\_k} opacity library and can use all the types of opacity sources available through it (correlated-$k$, cross sections, collision-induced absorptions, Rayleigh scattering, aerosol Mie scattering, etc.). The radiative transfer module has already been tested extensively for runaway greenhouse atmospheres\cite{2022A&A...658A..40C}.\\

\paragraph{Dry convective adjustment}
The dry convective adjustment scheme looks for convectively unstable regions in the atmosphere where the virtual potential temperature decreases upward.
The potential temperature and composition of unstable layers are fully mixed over a single timestep while conserving total mass, tracer mass, and enthalpy. \\

\paragraph{Moist convective adjustment and condensation}
Our scheme to handle moist convection is similar to the one used in previous models for runaway greenhouse atmospheres\cite{2013Natur.504..268L}. For water vapor dominated atmospheres as the ones treated here, the pseudo-adiabat reduces to the saturation vapor-pressure curve of pure steam.
Condensation of vapor can also happen in a layer in absence of convective processes, when there is diabatic cooling, for example. To account for this, at each timestep, this parametrization brings any supersaturated layer back to vapor equilibrium. Subsaturated layers can also evaporate condensates if they are present.
This saturation adjustment is performed iteratively until equilibrium conditions are found at constant moist enthalpy.  \\

\paragraph{Rains}
In the current version, all condensates precipitate instantaneously upon condensation, always leaving a cloud-free atmosphere. However, as we are interested in deep atmospheres where precipitations are unlikely to reach the surface (if it exists), we implemented a simple scheme to reevaporate falling precipitations. We start from the model top layer and collect all precipitations downward. Whenever an unsaturated layer is met, a fraction $f$ of the condensate that would need to be evaporated to saturate the layer is effectively evaporated. Precipitation are fully evaporated when they reach a layer where the temperature is above the boiling temperature of the falling species. All remaining precipitations when the surface layer is reached, if any, are added to the surface reservoir. 

In our study, precipitations are always reevaporated just below the saturated convection region and force a small, dry convective region there. This region experiences some level of time variability that is reminiscent of the results of 3D convection resolving models of hothouse Earth-like climates\cite{2021Natur.599...74S}.

\paragraph{Time integration}

Once we are able to compute heating rates in each layer ($\hat{H}\equiv\{H_\mathrm{n}\}_{\mathrm{n}=1,N_\mathrm{lay}}$; hatted symbols refer to vectors that denote the values of a given quantity in all the $N_\mathrm{lay}$ atmospheric layers at the same time), the temperature in each layer ($\hat{T}$) is integrated in time until thermal equilibrium is reached following
\begin{align}
c_p\frac{\partial \hat{T}}{\partial t} = \hat{H}(\hat{T}), \label{timeintegration}
\end{align}
where $c_p$ is the specific heat capacity of the gas.
In our case, only the equilibrium state of the atmosphere, where $\hat{H}(\hat{T})=\hat{0}$, is of interest. Therefore, a particular attention has been devoted to the development of schemes to accelerate the convergence toward such steady-states.\\

\noindent \textbf{Adaptive timestep.} 
As radiative transfer is usually the most expensive part of a 1D model, extra care has been taken to compute it as seldom as possible. As a first step in that direction, \texttt{Exo\_k} has an adaptive timestep that is based on the radiative timescale of the atmosphere. This radiative timescale is computed as follows. We start by saying that, around an arbitrary thermal state, say $\hat{T}_\mathbb{K}$, the radiative heating rates can be linearized through
\begin{align}
\hat{H}^\mathrm{rad}(\hat{T}_\mathbb{K}+\delta \hat{T}) = \hat{H}^\mathrm{rad}(\hat{T}_\mathbb{K}) + \mathbb{K} \cdot \delta \hat{T}, \label{jacobian}
\end{align}
where the kernel $\mathbb{K}$ is the Jacobian matrix of the heating rates (of dimension $N_\mathrm{lay} \times N_\mathrm{lay}$) and $\delta \hat{T}$ is a small temperature perturbation. 

This kernel matrix tells us how heating in a layer is related to a temperature change anywhere in our atmosphere. The biggest terms are the diagonal ones, which are usually negative. This shows that any layer whose temperature is increased will tend to emit more and cool in response. The terms directly above and below the diagonal -- the coupling terms between adjacent layers -- are usually positive but smaller in magnitude. Other off-diagonal terms, which are long range couplings, are much smaller, especially in optically thick atmospheres. 

Now, let us consider the radiative evolution of the system around radiative equilibrium --- meaning $\hat{H}^\mathrm{rad}(\hat{T}_\mathbb{K})=\hat{0}$. Combining the above equations yields the following evolution for the temperature perturbation:
\begin{align}
c_p\frac{\partial }{\partial t} \delta \hat{T}= \mathbb{K} \cdot \delta \hat{T}.
\end{align}

Keeping only the most important terms (the diagonal ones), the layers decouple and the solution reads
\begin{align}
\delta \hat{T} =  \delta \hat{T}_0 \cdot e^{- \hat{\mathrm{Diag}} (\mathbb{K}) t/c_p},
\end{align}
where $\hat{\mathrm{Diag}} (\mathbb{K})$ is the vector formed by all the diagonal terms in $\mathbb{K}$. So we see that in each layer, an initial perturbation, $\delta \hat{T}_0$, will be radiated away on a timescale equal to $\hat{\tau}= c_p/ \hat{\mathrm{Diag}} (\mathbb{K})$ that differs from layer to layer. In the baseline evolution, the smallest radiative timescale in our atmosphere, $\tau_\mathrm{rad}=\min(\hat{\tau})$, is used as timestep to ensure that the radiative evolution is well sampled throughout the atmosphere. We find that this condition is sufficient to ensure a stable evolution in most cases, but the user can always use smaller timesteps by specifying a factor $\alpha$ so that $\Delta t = \alpha \,\tau_\mathrm{rad}$.\\

\noindent \textbf{Using the Jacobian to compute fluxes.}
Another advantage of having computed the Jacobian is that, as long as the current state of the atmosphere is sufficiently close to the last state for which we computed $\mathbb{K}$, Eq.~\ref{jacobian} can be used to compute the radiative heating rates extremely rapidly through a simple matrix multiplication. 
In practice, we use this method as long as the maximum difference between the current temperature and $\hat{T}_\mathbb{K}$ is below some user defined threshold. Otherwise, the full radiative transfer is computed. This is especially efficient when the atmosphere is approaching equilibrium and requires a large number of small temperature increments to equilibrate its optically thick parts while the upper atmosphere requires small timesteps to remain stable. 
\paragraph{Convergence acceleration}
Despite the approach described above, the convergence time of an atmosphere remains prohibitive in many cases. This is usually due to the fact that the upper atmosphere has a short radiative timescale that requires small timesteps for stability while the deep atmosphere is very opaque and has a large thermal inertia. 

To circumvent these issues, the library offers several ways to accelerate the convergence toward an equilibrium state, as illustrated in \ref{fig:ED_Fig1_evol_acc}. It should be stressed, however, that the evolution trajectory followed by the atmosphere toward this state cannot be regarded as a temporal evolution, even tough we will refer to every evolution step as a \textit{timestep} for sake of brievity.

For each timestep, we first compute the heating rates for all the processes in the regular way described above over a duration $\Delta t$. We then identify \textit{radiative zones} as groups of adjacent layers where the heating rates of all the processes are zero, except for the radiation. A radiative zone can be composed of a single layer. The remaining layers of the atmosphere are grouped in stacks of adjacent layers that we will call \textit{convective zones} for convenience, even though the energy exchange in these zones might be due to formation or reevaporation of rains. \\
 
\noindent \textbf{Acceleration in radiative zones.}
We start by defining a base timescale, $\tau_\mathrm{b}$, as the smallest radiative timescale of a radiative layer in the atmosphere. Then, for any radiative layer $\mathrm{n}$, the heating rate is multiplied by $\tau_\mathrm{n}/\tau_\mathrm{b}$ before the timestepping is performed. 

In a purely radiative atmosphere this would be equivalent to advancing every layer independently using its own radiative timescale. In practice, this still allows deep radiative zones with very long timescales to converge in as many timesteps as their counterparts in the upper atmosphere. \\
 
\noindent \textbf{Acceleration in convective zones.} 
In convective zones, the layers in the zone are not independent as they directly exchange energy in a conservative way: except for radiation, energy that is taken in a part of the zone is redistributed in some other part. Using the scheme above in such zones completely alters the balance.

To understand why, let us consider a simple 2-layer convective zone where radiation heats the base layer and cools the upper one. When the layer becomes unstable, dry convection carries the surplus of heat from the bottom to the top layer. Equilibrium is reached when the convective energy flux equals both the net radiative heating of the base layers and the net radiative cooling of the top layer. Now, one can see that if the radiative heating rates are multiplied by two independent factors, our equilibrium solution is not an equilibrium anymore and the system will reach a new unphysical equilibrium.
\begin{figure*}[tb]
\centering
\includegraphics[width=\linewidth]{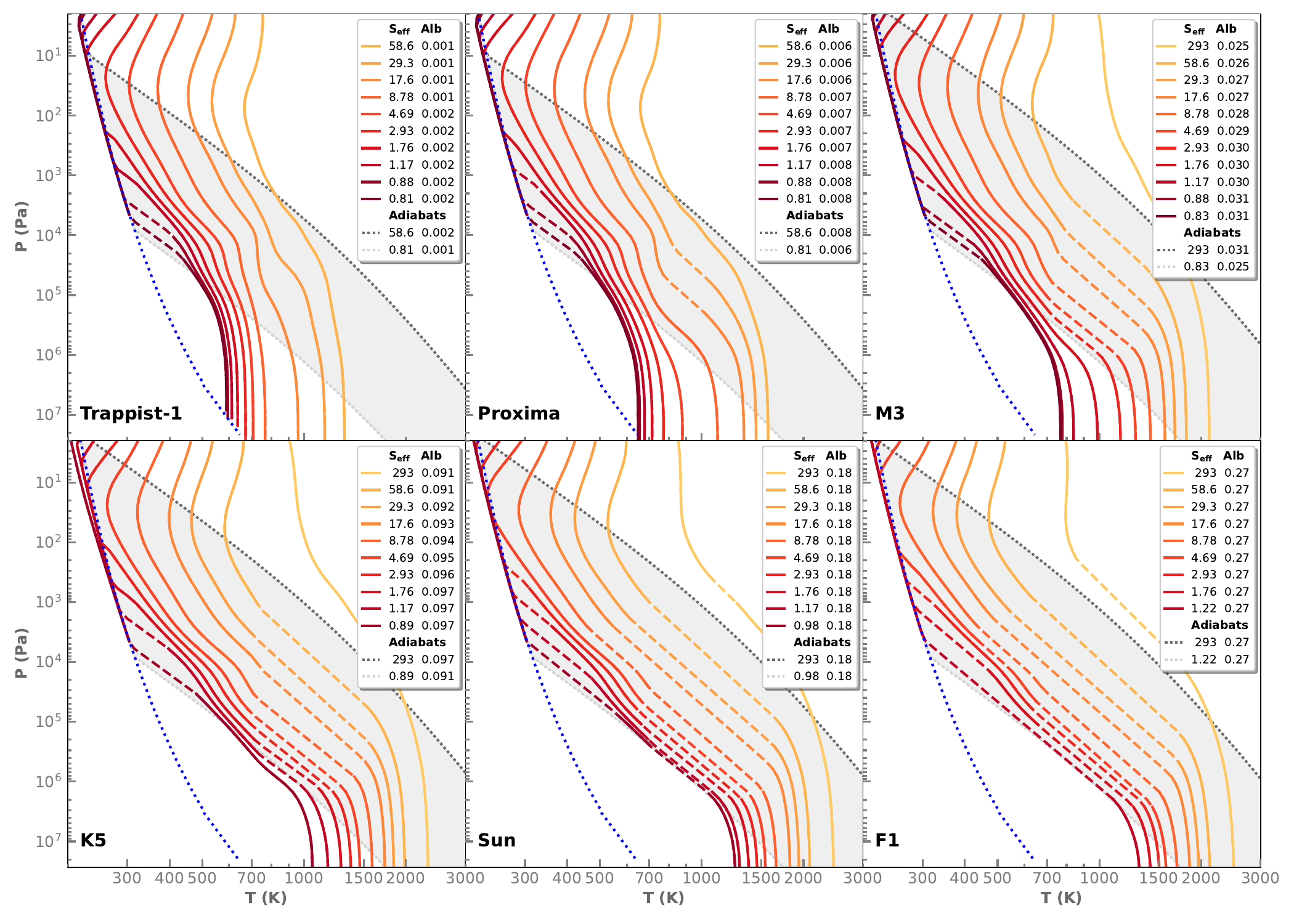}
\caption{\textbf{Steam atmosphere converged P-T profiles as a function of instellation and stellar type}. Dashed lines indicate dry convection. Adiabatic profiles satisfying top-of-the-atmosphere radiative balance for the minimum and maximum fluxes are in gray. }
\label{fig:ED_Fig2_all_profiles}
\end{figure*}
 
To avoid this problem, we treat an entire convective zone as one layer during the acceleration part. We compute the average radiative heating rate in the zone by computing the net radiative fluxes at the top and bottom of the convective zone and dividing it by its total mass. Then we compute a radiative timescale for the whole zone, $\tau_\mathrm{av}$. For the moment, we use the smallest radiative timescale of any layer inside the zone to be conservative. Finally, in addition to the heating rate computed in the initial step, each layer receives the same average radiative heating rate multiplied by $\tau_\mathrm{av}/\tau_\mathrm{rad}$.
  
\subsection{Simulations}
Unless indicated otherwise, simulations presented in this study were run for a pure H$_2$O atmosphere, a surface gravity of 1g and a surface pressure of 270~bars corresponding to vaporized Earth oceans. Our radiative transfer uses the H$_2$O linelists from HITRAN2016\cite{2017JQSRT.203....3G} and HITEMP2010\cite{rothman_hitemp_2010}. Above 700~K, we use HITEMP over the whole wavelength range. Below 700~K, we use HITRAN as much as possible because of considerable CPU time required to compute high-resolution spectra with HITEMP but mainly because the MT\_CKD H$_2$O-H$_2$O continuum parametrization (version 4.0.1\cite{2012RSPTA.370.2520M}) has been determined consistently with HITRAN (a continuum being basically the difference between the opacity measured and the opacity calculated with a given linelist when truncating the line profiles). Wavelength domains have been selected based on a comparison of opacities computed with both linelists\cite{2013NatGe...6..661G}: we use HITRAN at wavelengths above 3$\mu$m and HITEMP below 3$\mu$m. Opacities are treated with a correlated-k approach in 69 spectral bands from 0.28 to 1000~$\mu$m\cite{2021Natur.598..276T}. The model includes condensation and transport/evaporation of condensates but not their radiative effects. For the Sun, Trappist-1 (hereafter,T-1) and Proxima Centauri we use synthetic spectra\cite{2016A&A...596A..33R} while we approximate other stars (M3, K5 and F1) with a blackbody.\\
\ref{fig:ED_Fig2_all_profiles} shows converged profiles at different ISR for the Sun and other stars and \ref{fig:ED_Fig6_Psurf_Fint_EarlyVenus_T1} presents converged profiles for various surface pressures and internal heat fluxes.
\subsection{Comparison with a GCM} 
\begin{figure}[htb]
\centering
\includegraphics[width=\linewidth]{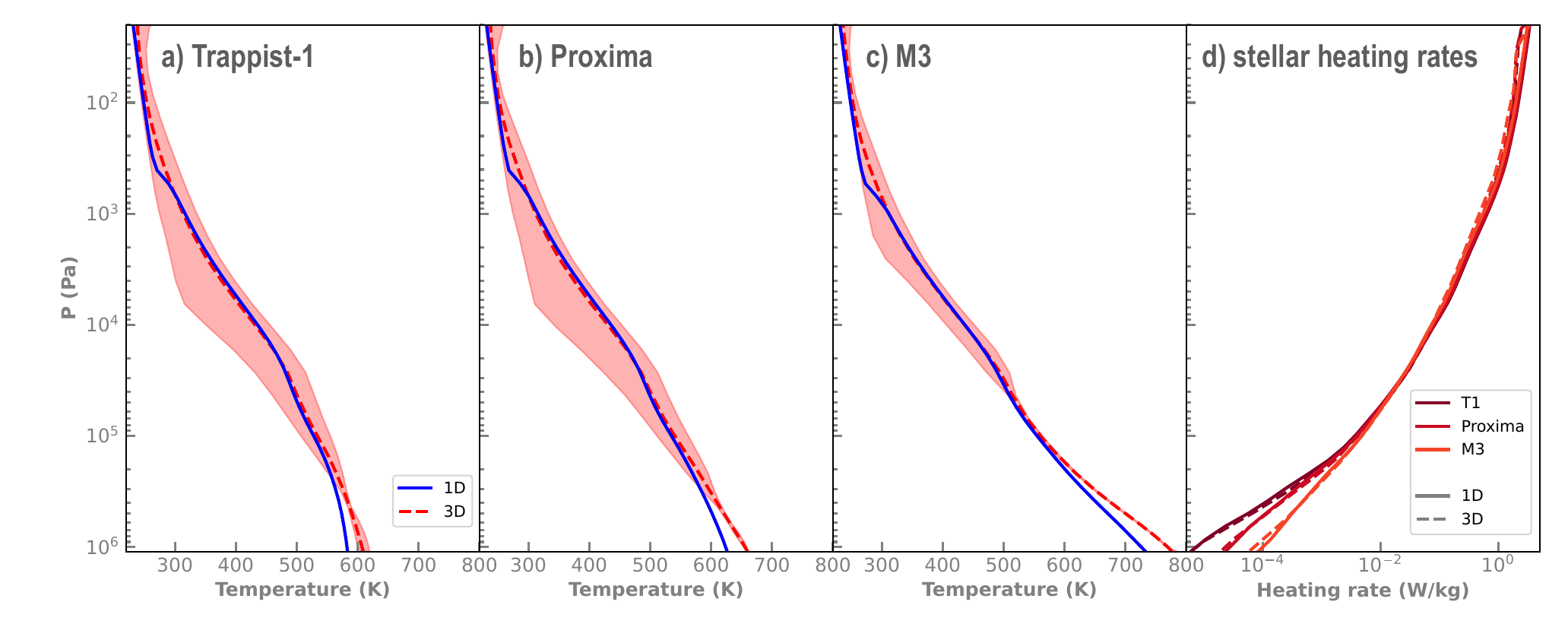}
\caption{\textbf{1D/3D comparison}. Panel a, b and c) Thermal profiles obtained with \texttt{Exo\_k} (solid blue lines) and the 3D Generic PCM (dashed red lines) for Trappist-1, Proxima and a M3 star. For the Generic PCM, the spatial and temporal average is shown as well as the range of variations (red area). Panel d) Stellar heating rates with \texttt{Exo\_k} (solid lines) and with the Generic PCM (dashed lines). The 10~bars atmosphere consists of 95\% of H$_2$O and 5\% of N$_2$ and the instellation is 500~Wm$^{-2}$ ($S_\textrm{eff}=1.42$) in all cases. The opacities used in both models are the same, differences come from circulation and cloud radiative effects.}
\label{fig:ED_Fig3_1Dvs3D}
\end{figure}

To test the results of \texttt{Exo\_k}, we compared them with those of the Generic PCM (Planetary Climate Model), previously known as the LMD Generic GCM\cite{2011ApJ...733L..48W,2021Natur.598..276T}. In addition to an existing simulation done for the Sun\cite{2021Natur.598..276T}, we performed new runs with the same parameters but colder stellar spectra and synchronous rotation. Convergence of 3D simulations was verified by starting from both hot and cold initial states\cite{2021Natur.598..276T}. Both 1D and 3D simulations were done for $p_\textrm{surf}=10$~bars as thicker atmospheres require prohibitively large CPU times in 3D. The comparison is presented in \ref{fig:ED_Fig3_1Dvs3D} and shows a good agreement. First of all, it confirms our main findings: the thermal structure \\
\noindent - does depend on the spectral distribution of the stellar radiation, \\
\noindent  - is radiative in the lower atmosphere, \\
\noindent - and more sub-adiabatic for cooler stars. \\
\texttt{Exo\_k} and Generic PCM are in close agreement for the stellar heating rates. This match is allowed by the absence of clouds on the dayside\cite{2021Natur.598..276T}. Thermal profiles show an excellent match at $p<10^5$~Pa, thanks to large-scale circulation maintaining a moist adiabat on the dayside at 10~mbar despite the absence of condensation. The Generic PCM is systematically hotter by $\sim 20$~K at the surface. This small departure is attributed to the greenhouse warming of nightside clouds\cite{2021Natur.598..276T}. Neither model finds convective layers at $p>10$~mbar.\\
Note that these 1D and 3D simulations were performed with the MT\_CKD 3.5 continuum. The 3D runs were done before the release of the 3.6 and 4.0.1 versions. While they are based on MT\_CKD~4.0.1 everywhere else in the study, the 1D runs shown in \ref{fig:ED_Fig3_1Dvs3D} are deliberately run with the same opacities as in the 3D model so that differences only result from 3D dynamics and clouds. By enabling the use of a 1D model, this comparison initiated the present study.
\subsection{Spectra}
\texttt{Exo\_k} can produce emission/reflection and transmission spectra like the synthetic spectra of the exoplanet Trappist-1b shown in figure~\ref{fig:Fig3_spectra_T1b}. We used the same spectral bands to compute the atmospheric structure and the spectra, as it is sufficient to display the effects described in the study. Spectral and profile calculations both account for the variation of gravity with altitude.  

\begin{figure}[tb]
\centering
\includegraphics[width=0.5\linewidth]{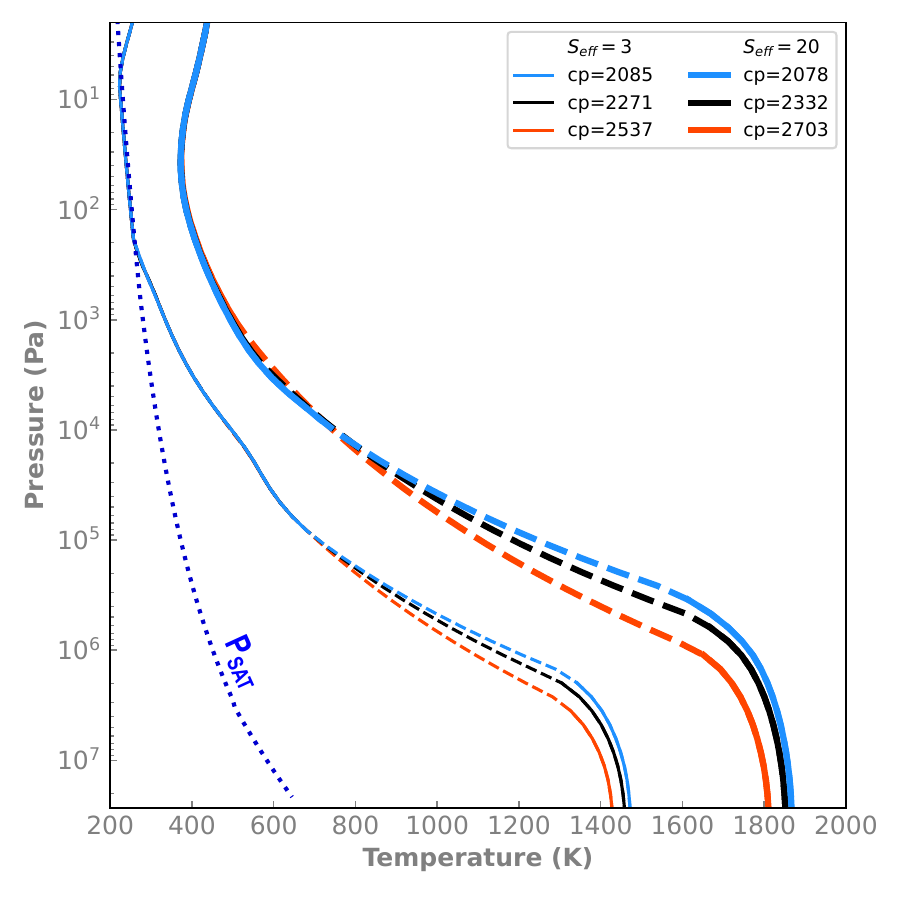}
\caption{\textbf{Influence of the $c_p$ value on the converged T profiles.} The black curves are the nominal P-T profiles, obtained with a $c_p$ set by iterations to its value at the mean temperature in the dry convection layers. To show the sensitivity to the $c_p$ value we used the lowest and highest temperatures ($T_{min}$ and $T_{max}$) found in these dry convective layers and computed the blue profile with $c_p(T_{min})$ and the red profile with $c_p(T_{max})$. Dashed lines indicate dry convection.}
\label{fig:ED_Fig4_profils_Sun_cp}
\end{figure}
\subsection{Sensitivity to opacity} \label{sub:opacity}
\begin{figure}[tb]
\centering
\includegraphics[width=0.8\linewidth]{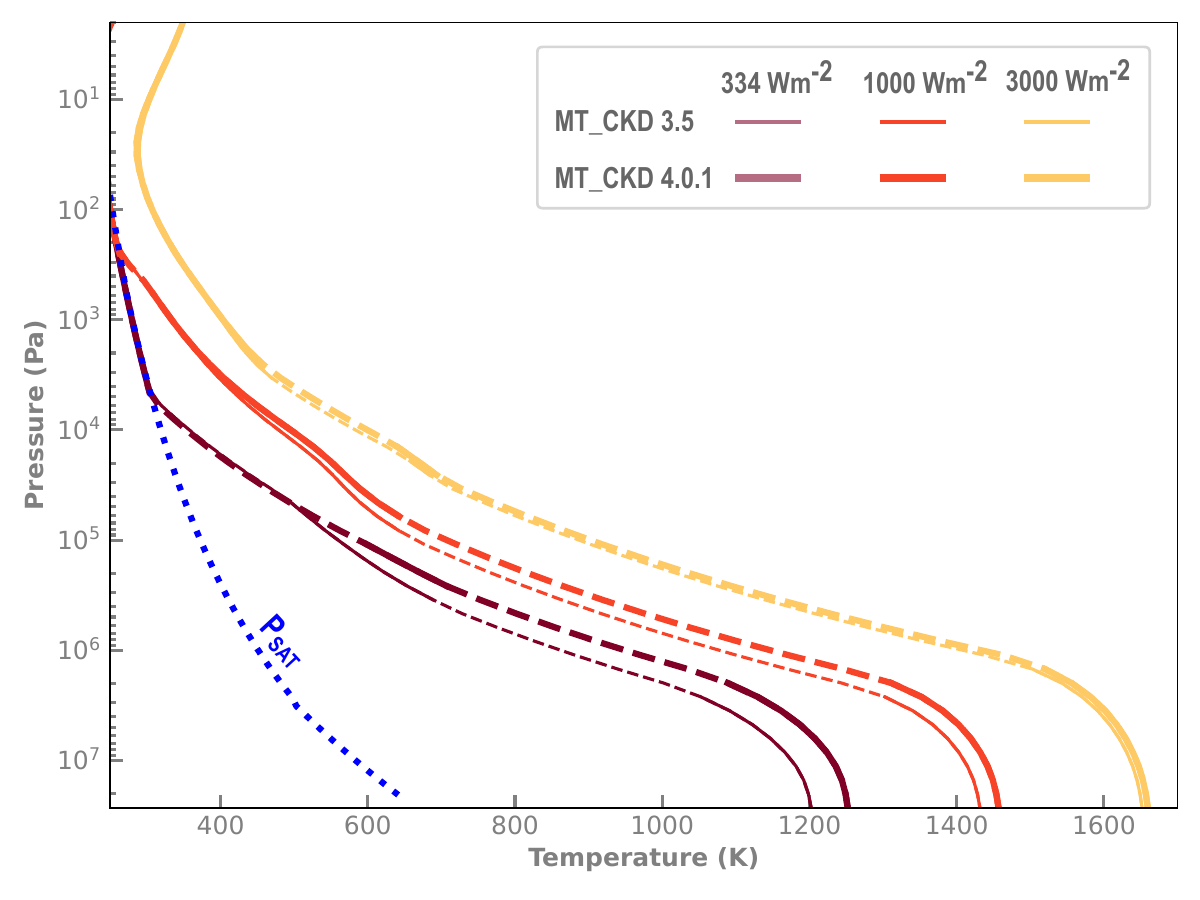}
\caption{\textbf{Sensitivity to a significant change in the H$_2$O-H$_2$O continuum.} Profiles are computed for a solar spectrum and three different insolations with versions 3.5 and 4.0.1 of the MT\_CKD continuum. Dashed lines indicate dry convection.}
\label{fig:ED_Fig5_continuum_compare}
\end{figure}

Radiative-convective profiles computed in this study depend on H$_2$O opacities that remain uncertain. In particular the radiative transport in optically thick layers is sensitive to far-wing collisional absorption in atmospheric windows. The continuum that controls this absorption is extrapolated from measurements made at temperatures and pressures far below those prevailing in the depths of a vaporized ocean. In \ref{fig:ED_Fig5_continuum_compare} we show how the calculated profiles are changed by the most recent update of the MT\_CKD H$_2$O self continuum. Compared to version 3.5, the temperature dependency in the version 4.0.1 was modified to account for new measurements and parametrizations\cite{2018AMT....11.2159L}. This comparison does not constitute in any way an estimate of the uncertainty on these profiles but they illustrate two points. First, despite orders of magnitude change in absorption in some windows, the strong temperature feedback on radiative diffusivity, variations of $T_\textrm{surf}$ do not exceed 50~K. Second, it however reminds us that new data will inevitably alter the quantitative results.
\subsection{Sensitivity to non-ideal behaviors of the equation of state}

In our simulations (either 1D or 3D), the gas specific heat capacity $c_p$ has the same value at all levels while it should depend on the temperature. The thermal profile is independent of $c_p$ in radiative regions, as well as in condensation layers (the gas being pure H$_2$O), but not in layers where dry convection keeps the thermal gradient at the dry adiabatic lapse rate $ \left. \frac{\partial \!\! \ \ln T}{\partial \!\! \ \ln p}\right\vert_{\mathrm{ad}}=R/\mu c_p $.  The value of $c_p$ must therefore be chosen consistently with the modeled case. When a deep dry-convection zone is present, we thus iteratively adjust $c_p$ during the convergence process to the value of the heat capacity at the mean temperature of the convective layers.

Although these convective regions are often limited in size, we did evaluate the uncertainty caused by using a constant $c_p$ by bracketing our nominal profile with profiles obtained with the lowest and highest values of $c_p$ found in the nominal dry-convective region. Results are shown in \ref{fig:ED_Fig4_profils_Sun_cp} for the solar case where convective zones are rather extended, so that it can be seen as a worst case scenario. The resulting changes on the equilibrium profiles are limited with a variation on $T_\textrm{surf}$ that does not exceed 50~K unless large internal heat fluxes ($>10$~Wm$^{-2}$) yield extended convective regions spanning a broad range of temperatures, which increases this sensitivity.

The library also uses the ideal gas law to compute the density of water vapor from pressure and temperature. This assumption is rather valid for high temperatures, but can underestimate the density by a significant factor near the critical point, which affects the computed altitudes  and, hence, the transit spectra. Although the library does not handle non-ideal equations of state in general, the altitudes used for the transmission spectra shown in figure~\ref{fig:Fig3_spectra_T1b} have been corrected for the non-ideal behavior of water vapor at high pressure. The largest effect was on the coldest Trappist-1b case that passes very close to the critical point and reduced the height of the atmosphere (transit radius) by less than 8\% (0.3\%).\\
\subsection{Effect of the internal heat flow and surface pressure} 
\begin{figure*}[hb]
\centering
\includegraphics[width=\linewidth]{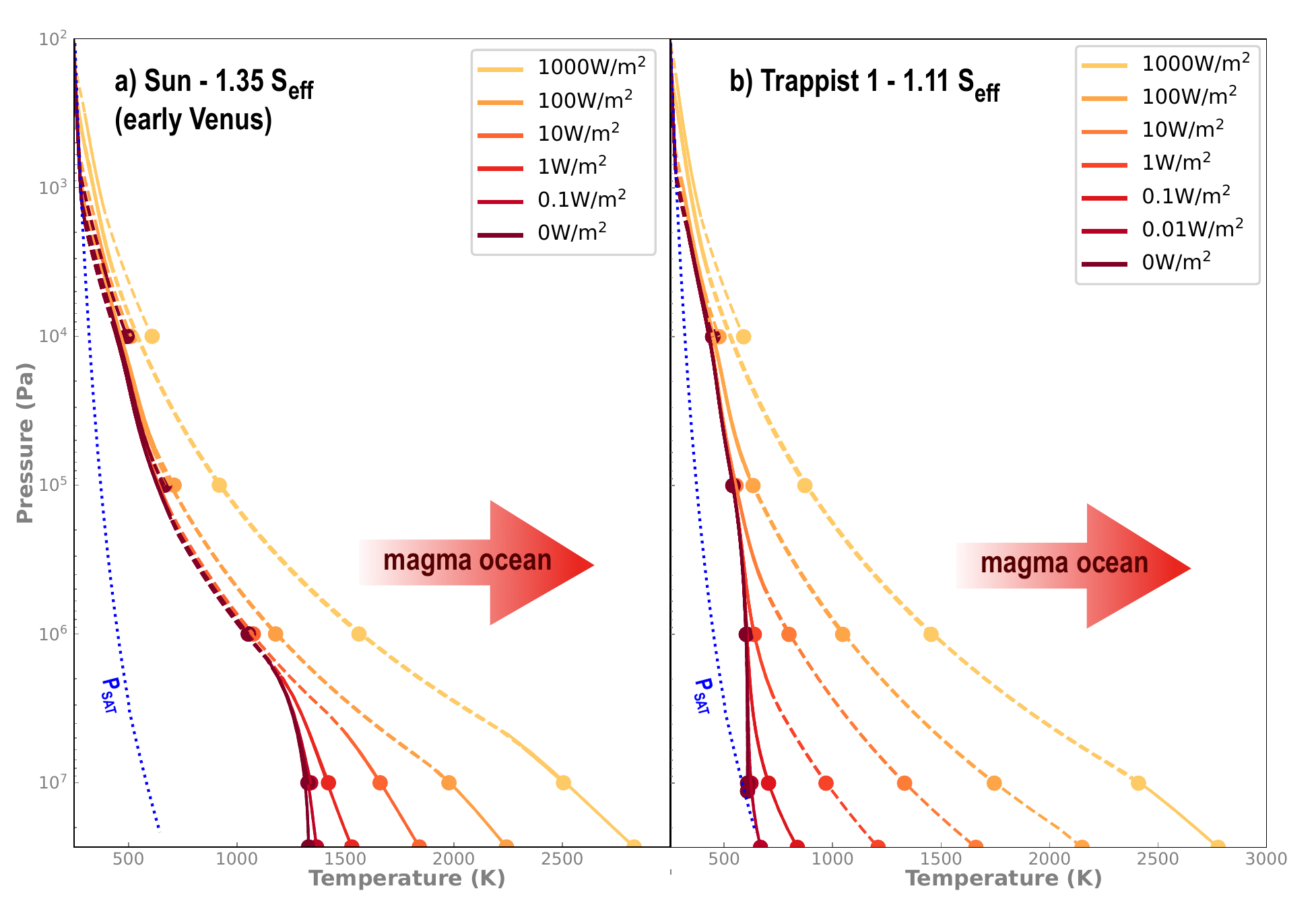}
\caption{\textbf{Influence of the surface pressure and internal heat flux.} a) Profiles for Venus gravity and insolation 4.5~Gyrs ago. b) Profiles for an Earth gravity with an ISR of 378~Wm$^{-2}$ and T-1 spectrum. At $\phi_\textrm{int}=0$ both cases have an OTR of 378~Wm$^{-2}$. Dashed lines indicate convective layers. Dots indicate the surface of each individual profile.}
\label{fig:ED_Fig6_Psurf_Fint_EarlyVenus_T1}
\end{figure*}

When the geothermal flux is comparable to or larger than the stellar flux reaching the surface, it contributes in heating the lower atmosphere. 
For a given heat flow and instellation, its effect thus strongly depends on the spectral type. Around the reddest dwarfs, the stellar flux at the surface can fall below the value of Earth present geothermal flux. As a consequence, the influence of an even low heat flux must be accounted for, while it can be neglected in the case of Sun-like irradiation, as shown in figure~\ref{fig:Fig2_runaway_exit}.
The surface pressure determines both the penetration of the stellar flux and the opacity of the deepest atmospheric layers, which control, with the temperature, the thermal gradient required to transport the surface heat flow upward. \ref{fig:ED_Fig6_Psurf_Fint_EarlyVenus_T1} shows how stellar spectrum, surface pressure and internal heat flux influence together surface temperature and the nature of the thermal profile (convective vs radiative). In some cases, increasing the geothermal flux can stabilize the lower atmosphere against convection thanks to the increase of radiative diffusivity with temperature. An early Venus with a steam atmosphere of more than 100~bars would not, for instance, develop a fully convective structure even for internal heat fluxes as high as 1000~Wm$^{-2}$ as its deepest layers would remain radiative. Under a pure steam atmosphere, the solidification of Venus surface would have occurred as soon as its internal heat flow decreases below a value that depends on the amount of water: 1000, 10 and 2 W/m2, for 10, 100 and 270 bars respectively.

\subsection{Composition of the atmosphere}
While the vaporisation of the Earth ocean by the increase of solar luminosity would produce, at least at first, a nearly pure steam atmosphere (with 0.3\% of N$_2$ and 0.0001~\% of CO$_2$ \cite{1988Icar...74..472K}), the atmosphere released by the accretion of a terrestrial planet or by a giant impact would be a mixture of outgassed volatiles. The composition of the resulting atmosphere depends on the composition of the (proto)planet, accreted bodies or large impactor, and on the distribution of the volatile compounds between the interior and the atmosphere. Although most studies of such atmospheres assume a mixture of H$_2$O and CO$_2$\cite{1986Natur.322..526M,1986Natur.319..303M,1988JAtS...45.3081A,2013JGRE..118.1155L,2015ApJ...806..216H}, 
the ratio between the two species is however poorly constrained, as well as the presence of other constituents like N$_2$, CO, H$_2$ \cite{2023E&PSL.60117894S} or sulfur-bearing compounds. The structure of atmospheres certainly varies a lot within this variety of compositions and conclusions derived for pure steam atmospheres cannot be generalized to post-accretion atmospheres. Exploring the thermal structure of these atmospheres goes far beyond the goal of the present study. However, the lesson learned from this work is that the convection hypothesis should at least be tested through the procedure described in \ref{subsec:method1}. Whenever \textit{inverse climate modeling} cannot be justified, \texttt{Exo\_k} can be used with appropriate opacities to compute a consistent structure.

\subsection{Stellar flux required to melt the surface} 
When the melt fraction exceeds the critical value of 0.4, a silicate mixture with a composition similar to that of the Earth mantle starts acquire the rheology of a fluid. In the low-pressure limit this fraction is reached at 1620~K \cite{Salvador_Samuel22}. \ref{tab:ED_Table_ISR_melt} gives the instellation that raises the surface to the \textit{melting} temperature as a function of the host-star type. Around Trappist-1, no planet is close enough to have its surface melted, including during the pre-main-sequence phase.  \\

\begin{center}
\begin{tabular}{@{}llll@{}}
\toprule
Star & $S_\textrm{eff}$ at solidus  & $S_\textrm{eff}$ for \textit{melting}$^{(*)}$ & $S_\textrm{eff}$ at liquidus \\
\midrule
Trappist 1    &  128    & \bf{339}  & -     \\
Proxima  & 24 & \bf{100} & 519   \\
M3    & 8.3   & \bf{24} & 167   \\
K5   & 3.5 & \bf{11} & 52 \\
Sun  & 2.1 & \bf{7.4} & 32 \\
\! \! \! \! +$\phi_\textrm{int}=0.09$~Wm$^{-2}$ & 1.8  & \bf{7.2} & 31 \\
F1 &   1.6  & \bf{6.2} & 28\\
\botrule
\multicolumn{3}{l}{$^{(*)}$ Melt fraction $> 0.4$, which implies a fluid rheology.} \\
\end{tabular}
\end{center}


\bmhead{ACKNOWLEDGEMENTS}
The authors sincerely thank Kevin Zahnle, Raymond T. Pierrehumbert, Robin Wordsworth and an anonymous referee for their particularly thorough and constructive review.
JL and FS acknowledge funding from the European Research Council (ERC) under the European Union’s Horizon 2020 research and innovation programme (grant agreement No. 679030/WHIPLASH), and from the french state: CNES, Programme National de Planétologie (PNP), the ANR (ANR-20-CE49-0009: SOUND), and in the framework of the Investments for the Future programme IdEx, Université de Bordeaux/RRI ORIGINS. \\
FS and MT acknowledge support from BELSPO BRAIN (project B2/212/PI/PORTAL). \\
MT acknowledges support from the Tremplin 2022 program of the Faculty of Science and Engineering of Sorbonne University. \\
The authors thank the Generic PCM team for the teamwork development and improvement of the model. \\
MT acknowledges the use of the High-Performance Computing (HPC) resources of Centre Informatique National de l’Enseignement
Superieur (CINES) under the allocation A0080110391 made by Grand Equipement National de Calcul Intensif
(GENCI), which was essential to compute the 3-D GCM simulations presented in this work. \\
EB and GC acknowledge support from the Swiss National Science Foundation grant 200021\_197176. Their work has been carried out within the framework of the NCCR PlanetS supported by the Swiss National Science Foundation under grants 51NF40\_182901 and 51NF40\_205606.  \\
This research has made use of NASA's Astrophysics Data System.\\

\bmhead{DATA AVAILABILITY}
Data generated by the atmospheric codes \texttt{Exo\_k} and \texttt{Generic PCM} and used in this study are available at \url{https://doi.org/10.5281/zenodo.6877001}.

 \bmhead{CODE AVAILABILITY}
 \texttt{Exo\_k} is an open-source software. A complete documentation on how to install and use it can be found at \url{http://perso.astrophy.u-bordeaux.fr/~jleconte/exo_k-doc/index.html}. \\
 The \texttt{Generic PCM} (Generic Global Climate Model, formerly known as LMDZ.generic) used in this work is the version 2528 that can be downloaded with documentation from the SVN repository at \url{https://svn.lmd.jussieu.fr/Planeto/trunk/LMDZ.GENERIC/}. More information and documentation are available at \url{http://www-planets.lmd.jussieu.fr}. 
 
\bmhead{AUTHOR CONTRIBUTIONS}
F.S. and J.L. laid the foundations for the study: F.S. wrote most of the paper, performed the 1D runs and made the figures -- with guidance from J.L, M.T, E.B. and G.C. -- while J.L. developed the code exo-k that made the study possible and wrote a significant part of the text, including exo-k description and manual. M.T. performed the 3D simulations, which were important to validate exo-k 1D profiles and pointed to radiative deep atmospheres. M.T., J.L. and G.C. worked on the spectroscopic data and formatted them for the study. All authors contributed to the response to the four reviewers and revisions of the article. 

\bmhead{COMPETING INTERESTS}
The authors declare that they have no competing financial interests.

\bmhead{CORRESPONDENCE}
Correspondence and requests for materials should be addressed to F.S. (franck.selsis@u-bordeaux.fr).

\bibliographystyle{sn-nature}
\bibliography{bibli_cool_runaway}


\end{document}